# The Mild Space Weather in Solar Cycle 24

Nat Gopalswamy[1], Sachiko Akiyama[1,2], Seiji Yashiro[1,2], Hong Xie[1,2] Pertti Mäkelä[1,2] and Grzegorz Michalek[3]

[1]NASA Goddard Space Flight Center
Code 671
Greenbelt, MD 20771
USA

[2]The Catholic University of America
620 Michigan Ave., N.E.
Washington, D.C., 20064
USA

[3]Astronomical Observatory of the Jagiellonian University
ul. Gołębia 24,
Krakow, Kraków County 31-007
Poland

**ABSTRACT**

The space weather is extremely mild during solar cycle 24: the number of major geomagnetic storms and high-energy solar energetic particle events are at the lowest since the dawn of the space age. Solar wind measurements at 1 AU using Wind and ACE instruments have shown that there is a significant drop in the density, magnetic field, total pressure, and Alfven speed in the inner heliosphere as a result of the low solar activity. The drop in large space weather events is disproportionately high because the number of energetic coronal mass ejections that cause these events has not decreased significantly. For example, the rate of halo CMEs, which is a good indicator of energetic CMEs, is similar to that in cycle 23, even though the sunspot number has declined by ~40%. The mild space weather seems to be a consequence of the anomalous expansion of CMEs due to the low ambient pressure in the heliosphere. The anomalous expansion results in the dilution of the magnetic contents of CMEs, so the geomagnetic storms are generally weak. CME-driven shocks propagating through the weak heliospheric field are less efficient in accelerating energetic particles, so the particles do not attain high energies. Finally, we would like to point out that extreme events such as the 2012 July 23 CMEs that occurred on the backside of the Sun and did not affect Earth except for a small proton event.

## 1. INTRODUCTION

Coronal mass ejections (CMEs) are a form of solar activity that cause large interplanetary disturbances including severe geomagnetic storms and large solar energetic particle events. CMEs originate in closed magnetic field regions, which appear as bipolar or multipolar regions in photospheric magnetograms. Many such magnetic regions contain sunspots, which served as the historical indicators of solar activity. There is usually a good correlation between sunspot number (SSN) and CME rate (Webb and Howard, 1994; Gopalswamy et al. 2010), but the correlation is not perfect because CMEs also originate from non-spot regions such as quiescent filament regions. Solar cycle 24 (SC 24) has been particularly interesting because it is the weakest in the Space Era and the space weather of solar origin has been extremely mild (Gopalswamy et al. 2014). However, the CME activity has not diminished proportionate to SSN, which declined by about 40% in SC24 (Gopalswamy et al. 2015). In this paper, we provide observational evidence that the mild space weather is caused by a combination of the weak state of the heliosphere influencing the properties of CMEs to make them less geoeffective.

## 2. SOLAR OBSERVATIONS IN CYCLES 23 and 24

Table 1 compares SSN, major geomagnetic storms (Dst ≤ -100 nT) and CME numbers over the corresponding epochs in SCs 23 and 24. It is clear that the number of major storms has declined by 76% in SC 24 relative to SC 23. This is a lot more than the 44% decline in SSN and 30% decline in fast and wide CMEs. We have used fast and wide CMEs because they are known to be responsible for causing major geomagnetic storms. In other words, there are appropriate CMEs, but they were not geoeffective. It is interesting to note that wide CMEs (width ≥60º) and halo CMEs (width =360º) have similar abundance in the two cycles, which gives a clue that there are large CMEs, yet they do not cause major storms.

Table 1. SSN, Major Magnetic Storms and CMEs in solar cycles 23 and 24

|  | Cycle23[a] | Cycle 24[b] | Ratio[c] | Change |
|---|---|---|---|---|
| SSN (61-mo average) | 68.27 | 38.45 | 0.56 | -44% |
| Storms (Dst ≤ -100 nT)[d] | 51 | 12 | 0.24 | -76% |
| CMEs (width ≥60º) | 1858 (32.6/mo) | 2205 (36.15/mo) | 1.11 | +11% |
| CMEs (width =360º) | 178 (2.99/mo) | 199 (3.06/mo) | 1.02 | +2% |
| Fast and Wide CMEs[e] | 189 (3.32/mo) | 142 (2.33/mo) | 0.70 | -30% |
| ≥C3.0 limb CMEs[f] | 273 (4.7/mo) | 214 (3.45/mo) | 0.73 | -27% |

[a]Over the same epoch as cycle 24; [b]1 December 2008 to 31 December 2013; some have different number of months, but same length in both cycles; [c]Ratio of cycle-24 rate to cycle-23 rate; [d]Storms until April 2014 included; [e]Speed ≥900 km/s and width ≥60º; [f]Limb CMEs associated with soft X-ray flares of size C3.0 or larger.

Figure 1 shows the continued decline in the number of geomagnetic storms relative to SSN as a function of time, binned over Carrington rotation (CR) periods. The decline is even deeper (~80%) given that the solar activity is already in the declining phase. SC 23 had another 32 major storms before it ended. It is unlikely that there will be comparable number of storms in the declining phase of SC 24. Since CMEs are responsible for major storms (except of occasional major storms caused by CIRs), some CME property must have drastically changed. Gopalswamy et al. (2014) found that for a given CME speed, SC 24 CMEs are significantly wider indicating an anomalous expansion of the CMEs in the coronagraph field of view. The confirmed this using measurements of plasma density, temperature, magnetic field at 1 AU. In particular, they found that the total pressure (magnetic + plasma) was significantly lower at 1 AU (by ~38%). Assuming that the lower pressure prevails near the Sun, the wider width can be explained by a faster expansion. For a given strength of the magnetic field ejected as a CME flux rope, one would expect that the field strength would have decreased significantly due to the expansion. After the initial expansion, the CME flux rope is supposed to evolve normally depending on the difference in pressure between the flux rope (magnetic cloud) and the ambient solar wind. The anomalous expansion may be one of the reasons why a larger number of CMEs with width

<30º are observed in this cycle. Many of these CMEs would have been too narrow to be counted in SC 23, but they are counted in SC 24, resulting in more counts. These small CMEs are not important for space weather because geomagnetic storms are caused by faster and wider CMEs (speed ≥900 km/s and width ≥60º) (Gopalswamy et al. 2010). In addition, the fast and wide CMEs need to come from close to the disk center in order that the CMEs directly impact Earth and produce geomagnetic storms.

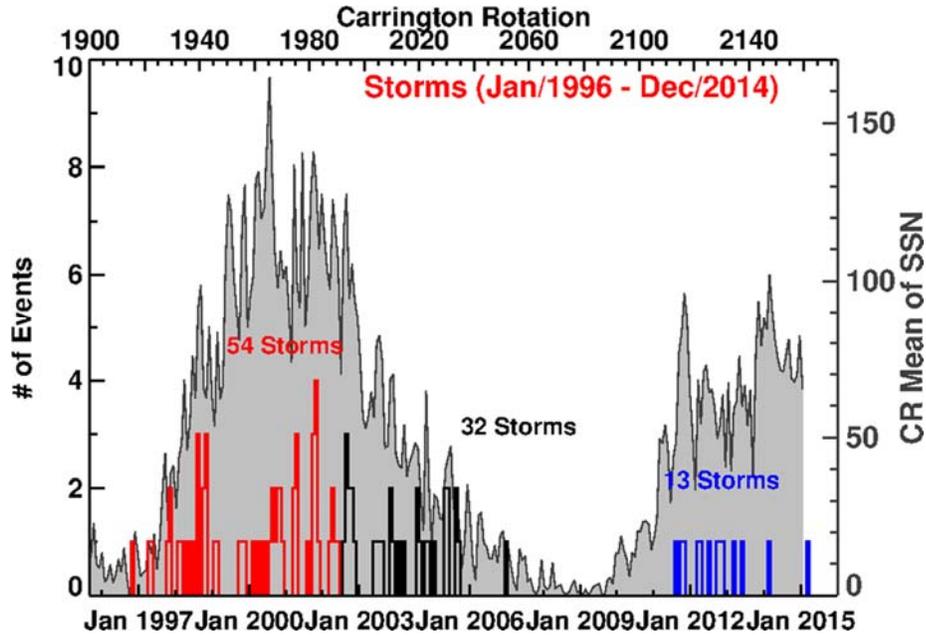

**Figure 1.** The cycle-24 storms and the ones in cycle 23 over the corresponding epoch (first 73 months) are distinguished by the blue and red lines, respectively. The black histogram shows the storms outside the study period. The number of storms in each interval is shown in the plot. The cycle-24 epoch shows also the storm on 2015 March 17, which is the most intense storm (Dst~-223 nT) in cycle 24 so far ((more on this event in the Discussion section). The grey distribution on the background represents the CR mean of SSN.

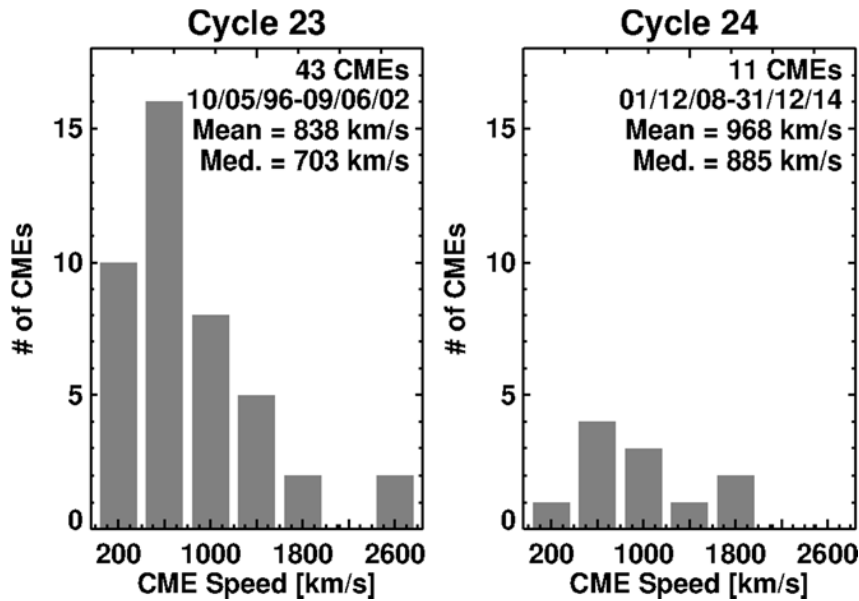

**Figure 2.** CME speed distribution in SCs 23 (left) and 24 (right) associated with major storms.

In order to see if there is any difference in the CMEs that caused major geomagnetic storms in SC 24, we have shown the speed distribution of CMEs associated with major storms SCs 24 and 23 in Fig.2. The shape of the distributions look similar and the average CME speed is slightly higher in SC 24. There are 11 CMEs in SC 24 until the end of 2014 because one of the major storms that occurred on 2013 June 1 at 9:00 UT (Dst = -108 nT) was due to a CIR. The 11 CME-related storms had Dst in the range -107 to -147 nT. However, the range of Dst in SC 24 was very narrow (> -150 nT) compared to -387 nT in SC 23. In other words, the CMEs of SC 24 have to be even faster to produce the same-size or smaller geomagnetic storms. The intensity of geomagnetic storms is primarily determined by the speed and the southward magnetic field component of the interplanetary magnetic field. Therefore, it depends how the CMEs evolve in the IP medium before arriving at 1 AU. Two modifications are expected: (i) the ambient solar wind speed has also declined in SC 24, so the drag force acting on CMEs is larger and hence the CMEs are likely slow down more; (ii) the magnetic content of the CMEs is expected decrease (lower field strength). These two can be verified from solar wind observations of CMEs in the two cycles.

## 3.2 SOLAR WIND OBSERVATIONS AND GEOMAGNETIC STORMS

Gopalswamy et al. (2008) found an empirical relation, Dst = -0.01VBz -25 nT, where V is the speed (in km/s) of the magnetic cloud (MC) at 1 AU and Bz (in nT) is the magnitude of the southward component of the MC magnetic field (see also Echer et al. 2005). Magnetic clouds are the CME counterparts at 1 AU that are supposed to be heading directly toward the observer with a flux rope structure (see e.g., Burlaga et al. 1981). There are other non-cloud structures at 1 AU, but we consider only MCs in this work (see also Dasso et al. 2012).

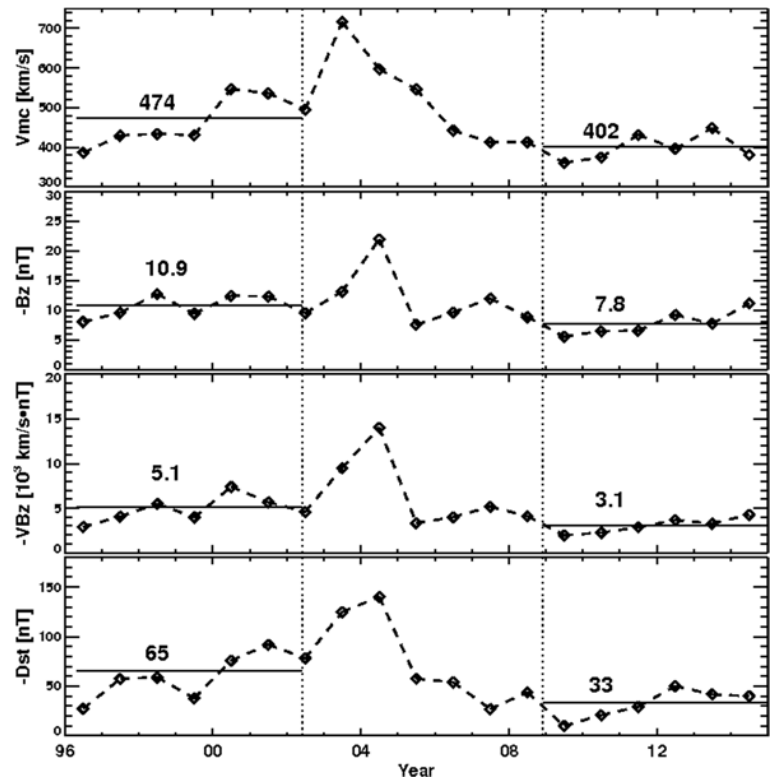

Figure 3. Annual averages of MC properties from top to bottom: speed, Bz, VBz, and Dst. SC 24 starts from December 2008 (to the right of the second vertical dotted line); the corresponding epoch in SC 23 is to the left of the first vertical dotted line. The horizontal bars are 73-month averages.

We identified 68 MCs in SC 23 and 65 in SC 24, which are similar in number, a property shared by fast and wide CMEs or frontside halo CMEs in the two cycles (Gopalswamy et al. 2015). We compiled the speed and field strength of the MCs and the associated peak Dst index. Figure 3 shows the annual averages of the speed ($V_{MC}$), Bz, and VBz of the MCs and the associated Dst. Clearly all quantities declined in SC 24 compared to the corresponding epoch in cycle 23. Gopalswamy et al. (2014) found that the average speeds of CMEs did not differ significantly between the two cycles (658 km/s in SC 23 and 688 km/s in SC 24), but the widths did. The 1-AU average speeds in Fig. 3 show a significant decline (15%), although for not the same set of CMEs. Thus the slowdown of SC 24 CMEs is more pronounced, most likely due to the enhanced drag in the weaker cycle. Bz declined by 28%. The MC speed – MC Bz product VBz declined by 39%. The decline in average Dst is by 49%, similar to the decline in VBz as expected. Thus lower geoeffectiveness in SC 24 seems to be determined by the lower VBz in MCs. The SC 24 empirical relationship Dst = -0.017 VBz +16 nT obtained for the 65 MCs is not very different from the corresponding relationship for the SC 23 MCs: Dst = -0.013 VBz -6 nT (and the one for the full SC 23 from Gopalswamy et al. 2008). Figure 4 emphasizes this point using the distributions of VBz and Dst for the entire period of 73 months in each cycle. The average values are the same as in Fig. 3, but we can see the ranges of VBz and Dst are narrow in SC 24. In SC 24, only a third of the MCs were geoeffective (Dst <-50 nT), while more than half of the MCs were geoeffective in SC 23.

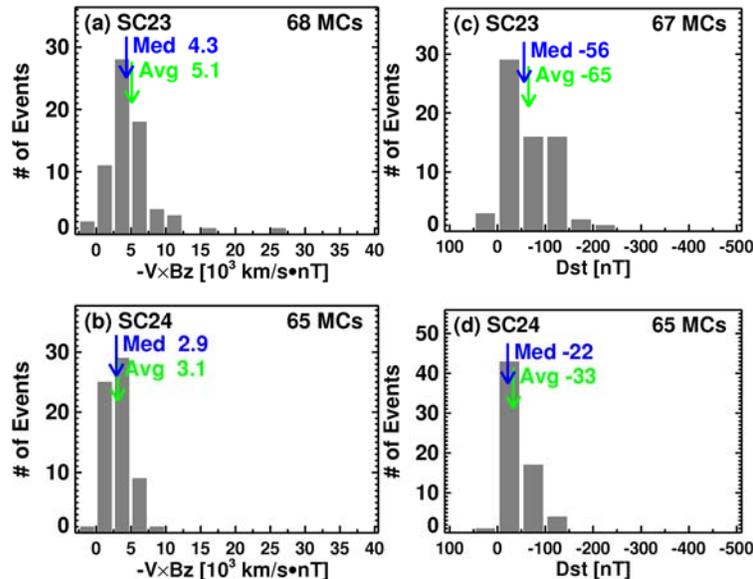

Figure 4. The product VBz of magnetic clouds (left) and the associated peak Dst index (right) for SC 23 (left) and 24 (right). The average (Avg) and median (Med) values are shown on the plots.

## 4. DISCUSSION

We have shown the low geomagnetic activity in SC 24 can be traced to the low VBz in the interplanetary counterparts of CMEs. The storms generally were of low intensity and infrequent, even though the number of fast and wide CMEs declined only slightly in this cycle. We suggest that the dilution of the magnetic content of CMEs near the Sun due to the increased expansion of CMEs in SC 24 seems to be the reason, coupled with the enhanced drag on CMEs due to the diminished solar wind speed in this cycle.

We noted in the beginning that a super geomagnetic storm occurred on March 17, 2015 with a Dst of -223 nT. The activity is already in the declining phase, and it is not uncommon for such intense storms occur in the declining phase (e.g., the Halloween 2003 storms, see Gopalswamy et al. 2005a). This storm was caused by an Earth-directed CME that erupted on March 15 2015 from close to the disk center (heliographic coordinates S18W39) with a speed of 1120 km/s in the sky plane. This can be converted to an earthward speed of ~1025 km/s based on the source location on the Sun. The CME arrives as a magnetic cloud on March 17 at 13:38 UT with a large southward field in the sheath (-15 nT) and in the front of the cloud (-25 nT). There are Dst depressions corresponding to the sheath and cloud portions, with the deepest minimum Dst occurring at 23:00 UT on March 17. The speed of the MC was ~600 km/s, considerably smaller than the white-light CME speed of 1025 km/s. However, the speed decline is similar to the average slow down noted before (688 km/s to 402 km/s or 42% on the average). The slowdown of the March 15 CME is by nearly the same extent: 1025 km/s to 600 km/s or 41%.

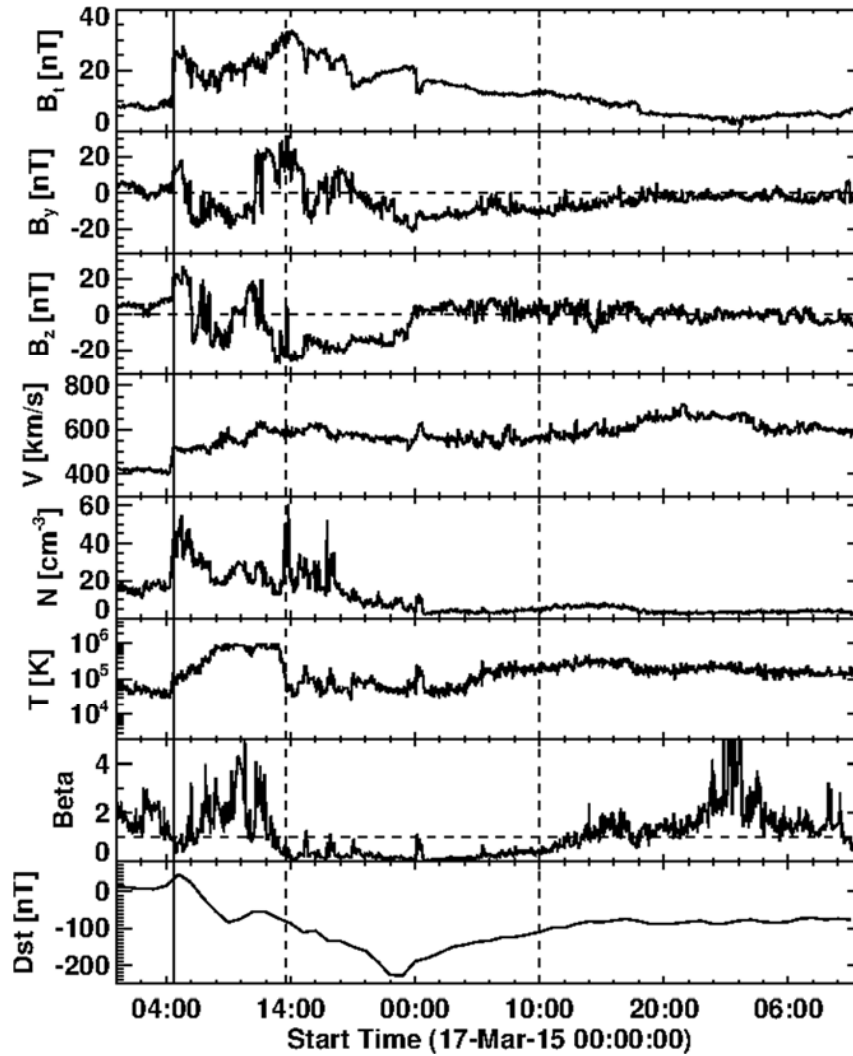

Figure 5. Time profiles of plasma parameters, magnetic field components (OMNI 1-minute data) and the Kyoto Dst index (1-hour time resolution) during 17-19 March 2015. The dashed vertical lines show the approximate boundaries of the interplanetary CME (17/13:38 UT – 18/10:00 UT) that caused the most intense geomagnetic storm (Dst~-223 nT) in cycle 24 so far. The associated shock (vertical solid line) is seen at ~17/04:34 UT.

The geomagnetic response is also consistent with the empirical relationship we noted before for SC 24: Dst = -0.017VBz +16 nT. Plugging in V = 600 km/s, Bz = -25 nT, we get Dst = -239 nT, nearly the same as the observed value. Without the anomalous expansion and the increased interplanetary drag, this storm would have been much bigger. If we use the average drop of 39% for VBz between cycles 23 and 24, we see that a storm like this would have had a VBz of 24600 km/s.nT and the Dst would have been -402 nT, similar to the largest storm of solar cycle 23 (Gopalswamy et al. 2005b).

We note that there was an extreme event that happened on the backside of the Sun on 2012 July 23 (Baker et al. 2013). It was observed as a large backside CME by SOHO, so it did not affect Earth, except for a small energetic particle event. However, the monstrous CME was heading toward STEREO-Ahead spacecraft and reached there in about 19 hours. The MC had a speed of 1500 km/s and Bz = -52 nT. If it were Earth-directed, the Dst index would have plunged to -1310 nT, similar to the historical Carrington event. If the same event had occurred in cycle 23, the Dst would have fallen below -2100 nT. Such extreme events may occur at any time of the solar cycle, but what we have described is the average behavior of the cycle and how the geoeffectiveness of CMEs can be drastically different.

The total pressure in the heliosphere that causes the anomalous expansion of CMEs is accompanied by a reduction in the heliospheric magnetic field. This affects the other space weather aspect: energetic particle events from the Sun. The highest energy particles travel down to the troposphere, produce secondary neutrons in air showers that are observed by ground based neutron monitors. These are known as ground level enhancement (GLE) events. Typically a dozen such GLE events occur in each solar cycle, however, SC 24 had only 2 GLE events. Compared to the 9 events in SC 23 over the same epoch, the decline is by ~78%, similar to the major geomagnetic storms. Although there are shocks and type II radio bursts in the interplanetary medium indicating particle acceleration, the particles do not seem to be accelerated to very high energies. Gopalswamy et al. (2014) attributed the paucity of GLEs to the diminished acceleration efficiency of shocks because the efficiency is proportional to the ambient field strength.

## 5. CONCLUSIONS

The main conclusion of this study is that the reduced geomagnetic activity of SC 24 can be directly attributed to the reduced field strength and the speed of magnetic clouds arriving at Earth. This is because the minimum value of Dst is proportional to the product of the magnetic cloud speed and the magnitude of the southward component of the cloud field (VBz). We attribute the reduction in Bz to the anomalous expansion of CMEs near the Sun. The reduced cloud speed at 1 AU can be attributed to the enhanced drag in SC 24 due to diminished solar wind speed and increased CME size. Since the drag force is proportional to the square of the speed difference between the CME and the solar wind, a slower wind will result in a larger drag. Similarly the drag is also proportional to CME cross sectional area, so wider CME in cycle 24 provides a larger area and increases the drag.

**ACKNOWLEDGEMENTS**
We acknowledge data use from the OMNI data base at NASA GSFC and the Dst index from the World Data Center in Kyoto. SOHO is a project of international collaboration between ESA and NASA. STEREO is a mission in NASA's Solar Terrestrial Probes program. The work of NG, SY, SA was supported by NASA/LWS program. PM was partially supported by NSF grant AGS-1358274 and NASA grant NNX15AB77G. HX was partially supported by NASA grant NNX15AB70G. GM was supported by NCN through the grant UMO-2013/09/B/ST9/00034.